\documentclass[aip,rbi,reprint,graphicx]{revtex4-1} 
\usepackage[dvips]{graphicx}
\usepackage{color}

\begin{document}


\title{Sub-Kelvin refrigeration with dry-coolers on a rotating system}



\author{S.~Oguri}
\email[]{Author to whom correspondence should be addressed.
Electronic mail: shugo@post.kek.jp}
\affiliation{Institute of Particle and Nuclear Studies, High Energy Accelerator Research Organization~(KEK), Oho, Tsukuba, Ibaraki 305-0801 Japan}
\author{H.~Ishitsuka}
\affiliation{Department of Particle and Nuclear Physics, School of High Energy Accelerator Science, The Graduate University for Advanced Studies (SOKENDAI), Shonan Village, Hayama, Kanagawa 240-0193 Japan}
\author{J.~Choi}
\affiliation{Korea University, Anam-dong Seongbuk-gu, Seoul 136-713
Republic of Korea}
\author{M.~Kawai}
\affiliation{Institute of Particle and Nuclear Studies, High Energy Accelerator Research Organization~(KEK), Oho, Tsukuba, Ibaraki 305-0801 Japan}
\author{O.~Tajima}
\affiliation{Institute of Particle and Nuclear Studies, High Energy Accelerator Research Organization~(KEK), Oho, Tsukuba, Ibaraki 305-0801 Japan}
\affiliation{Department of Particle and Nuclear Physics, School of High Energy Accelerator Science, The Graduate University for Advanced Studies (SOKENDAI), Shonan Village, Hayama, Kanagawa 240-0193 Japan}


\date{\today}

\begin{abstract}
 We developed a cryogenic system on a rotating table
 that achieves sub-Kelvin conditions.
 The cryogenic system consists of a helium sorption cooler
 and a pulse tube cooler in a cryostat mounted on a rotating table.
 Two rotary-joint connectors for electricity and helium gas circulation
 enable the coolers to be operated and maintained with ease.
 We performed cool-down tests
 under a condition of continuous rotation at 20~rpm.
 We obtained a temperature of 0.23~K
 with a holding time of more than 24 hours,
 thus complying with catalog specifications.
 We monitored the system's performance for four weeks;
 two weeks with and without rotation.
 A few-percent difference in conditions was observed between these two states.
 Most applications can tolerate such a slight difference.
 The technology developed is useful for various scientific applications
 requiring sub-Kelvin conditions on rotating platforms.
\end{abstract}

\pacs{%
07.20.Mc, 
%
07.57.Kp, 
%
11.30.Cp, 
%
14.80.Va, 
%
84.40.-x, 
%
95.85.Bh, 
%
98.80.Bp, 
98.80.Es  
}


\maketitle 



Making a low temperature condition on a rotating platform is a key
technology in a number of science experiments, e.g., tests of Lorentz
invariance~\cite{PhysRevLett.95.040404,PhysRevA.71.050101}, searches for
axions~\cite{PhysRevD.78.032006}, and radio telescopes for astronomy and
cosmology~\cite{GroundBIRD}.
In a previous publication~\cite{prev_paper}, we described
a cryocooling system on a rotating table;
the lowest temperature achieved was 7~K
using a two-stage Gifford-McMahon cryocooler.
In this paper, we report a lowering of the lowest temperature to 0.23~K.
This advancement allows superconducting devices to be used in the system.
In particular, attaining temperatures below 0.3~K is
an important milestone in regard to precise measurements
of the cosmic microwave background radiation~\cite{GroundBIRD}.

\begin{figure}
 \includegraphics[width=\linewidth]{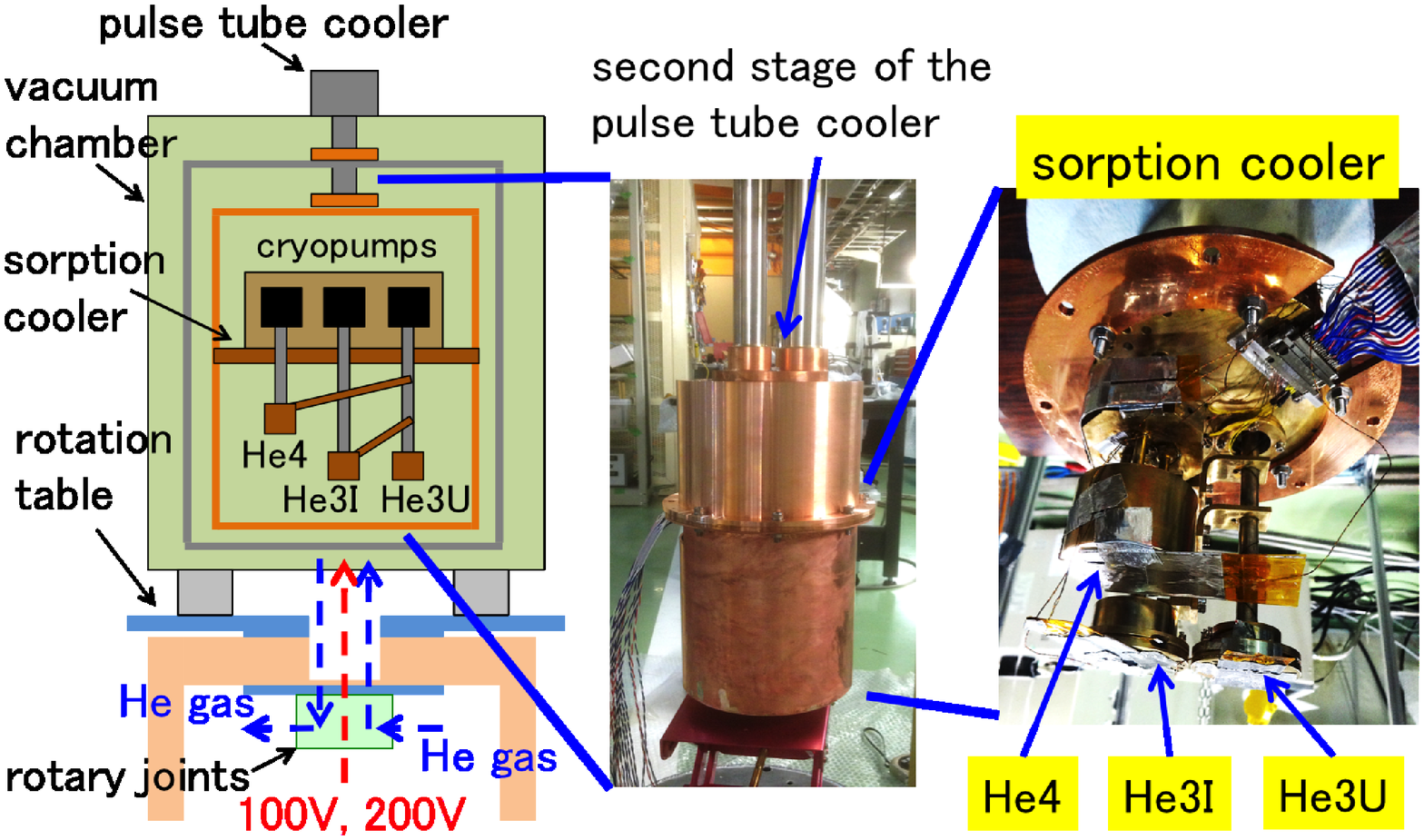}
 \caption{System schematics.
 The helium sorption cooler is set at the second stage of the cryocooler.
 The sorption cooler was surrounded by a copper shield
 which was maintained at $\sim$~4~K.}
 \label{fig:setup}
\end{figure}

The sub-Kelvin refrigeration system consists of
a pulse tube cryocooler\cite{refPTC}
and a helium sorption cooler\cite{refsorp2}.
A schematic of the cryogenic system with rotation table
is shown in Fig.~\ref{fig:setup}.
The cryostat on the table holds a two-stage pulse tube cryocooler
(PT415-RM, Cryomech, Inc., Syracuse, NY, USA);
the cooling power for the first stage is 36~W at 45~K
and for the second stage is 1.35~W at 4.2~K.
A series of two rotary joints maintains
the simultaneous circulation of electricity and helium gas
between the cryocooler and a compressor positioned on the ground.
Further details about the operation of the cryocooler are described in
our previous publication~\cite{prev_paper}.

For ease in handling the helium hoses,
we used 1/2''-diameter hoses on the rotating table
although the cryocooler specifications require 3/4''-diameter hoses.
Hence, we changed the diameters of the input/output ports
for the gas joint and of the ports of the motor head for the cryocooler
by using additional connectors between each port.
Differences in conductance produced a difference in performance
of the cryocooler%
\footnote{The readjustment of the motor frequency is necessary for the
maintenance of the performance whereas we did not readjust it.}.
However, this difference was small enough for tests to be performed
as the expected cooling power loss was at most $\approx$~0.3~W
as determined by heat capacity measurements.


For further cooling below 4~K, we used a three-stage helium sorption
cooler (CRC\_GL10, Chase Research Cryogenics Ltd., Sheffield, UK)
set at the bottom of the second stage of the cryocooler
(Fig.~\ref{fig:setup}).
Its base plate temperature was kept at 3.8~K.
Each stage has a tank which stores liquid helium and a cryopump
which adsorbs helium vapors.
Using the cryopump, each stage reaches its desired cold temperature
by vaporization cooling.
As the three stages contain sequentially $^4$He, $^3$He, and $^3$He,
we labeled the coldest place in each stage (and each cryopump)
``He4'' head (pump), ``He3I'' head (pump), and ``He3U'' head (pump),
respectively%
.
The stages are thermally connected in series; the temperatures attained
decrease in the order He4 head, He3I head, and He3U head.
The temperatures specified in the catalog for each head are approximately
0.25~K (He3U head), 0.35~K (He3I head), and 1.0~K (He4 head) under
loading conditions of 3~$\mu$W, 40~$\mu$W, 200~$\mu$W, respectively.
The holding time for the He4 head is more than 24 hours
(the other two heads have longer holding times under these conditions).
We confirmed these specifications in advance using a different setup.


Each cryopump was controlled via two heat switches:
the ``cold'' and ``warm'' switches.
The cold switch maintains the cryopump, where helium vapor is adsorbed,
in a cold state ($\sim$4~K).
The warm switch \footnote{This is just a resistance heater in fact.}
raises the cryopump temperature (up to 55~K), i.e., it increases pressure in the tank.
Liquid helium accumulates in the tank at conditions above vapor pressure.
That is, each cold head reverts to the cold state when the cold switch is turned on.
In contrast, each tank is refilled with liquid helium when the warm switch is turned on.
For all three stages, we call this refilling cycle of the liquid helium ``recharge''.

\begin{figure}
\includegraphics[width=0.9\linewidth]{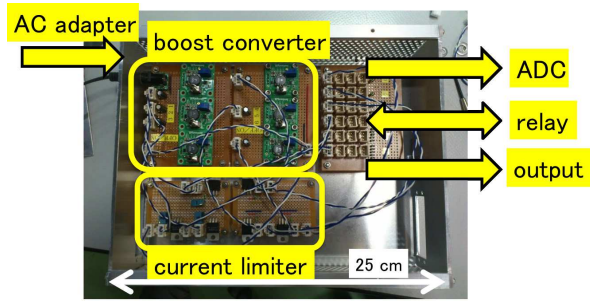}
 \caption{Photo of the compact size switch controller necessary
 for the operation of the sorption cooler.
 Six boost converters and six current limiters are used
 in the controller.}
 \label{fig:pic_circuit}
\end{figure}

For the control of each cold switch, we need a power supply  of 3.5~V DC.
To operate the warm switches at the He4, He3I, and He3U pumps,
we also need a power supply of 20~V, 15~V, and 10~V DC, respectively.
Recharging requires controlling these switches.
To save space on the table, we developed a compact size switch controller.
The controller consists of an AC adapter, six boost converters,
and six current limiters.
Supply of the DC 5~V power was performed using an AC adapter
from the single-phase AC 100-V power supply,
which is connected via the electric rotary joint~\cite{prev_paper}.
The boost converters supply the correct DC voltage to each switch.
The current limiter consists of a 3-terminal regulator
(LM317, Texas Instruments Inc., Dallas, Texas, USA).
Each current into the switch is monitored using an analog-to-digital
converter (ADC-16, Pico Technology, Cambridgeshire, UK).
Figure~\ref{fig:pic_circuit} shows a photograph of the circuit
for the boost converters and the current limiters.
The output power for each switch was controlled by built-in relays of
a temperature monitor
(Model 218, Lake Shore Cryotronics, Inc., Westerville, Ohio, USA).
This monitor was also controlled by a computer via its serial interface.


There were ten thermometers on the sorption cooler.
A Cernox$^{\rm TM}$ resistance temperature sensor
(Lake Shore Cryotronics, Inc., Westerville, Ohio, USA)
was mounted on the He3U head.
A ruthenium oxide resistance sensor was mounted on both the He3I and He4 heads.
These three resistance sensors were measured using AC resistance
bridges built-in on a cryogenic temperature controller (Model 44,
Cryogenic Control Systems, INC., Rancho Santa Fe, CA, USA).
The remaining seven thermometers were silicon diode sensors;
they measured temperatures of each heat switch and at an intermediate
location between the He4 head and the He4 pump.
They were monitored using the temperature monitor
which was used for the heat switch control.
All data were recorded and stored on the computer at 10~seconds interval.

\begin{figure}
 \includegraphics[width = 0.9\linewidth]{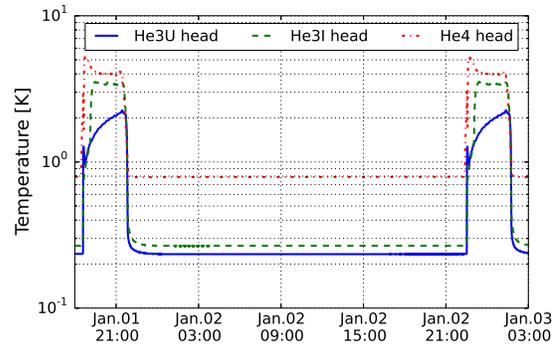}
 \caption{
 An example of refrigeration cycles for the sorption cooler
 on the rotation table at 20~rpm.
 Variation of temperatures at each cold stage is shown in this plot.
 We confirmed that the lowest temperature and holding time
 were consistent with specifications of the sorption cooler.}
 \label{fig:one-cycle}
\end{figure}

We operated the sub-Kelvin cooling system on the rotating table at 20~rpm.
Figure~\ref{fig:one-cycle} shows temperature trends of each head in the
interval between each recharge cycle.
The temperature of each head fell below
234~mK (He3U), 267~mK (He3I), and 787~mK (He4), respectively.
The time interval between two recharge cycles was more than 24 hours.
This is consistent with the specifications of the sorption cooler.

Control during recharging was made easy using software based on Python scripts;
the program executed temperature monitoring and control switching.
An ``auto-recharge'' program was used to monitor long-term trends.
Recharging starts when liquid helium in the He4 head is exhausted;
the timing is synchronized
when the temperature of the He4 head rises above 1~K.
During liquid helium refilling, all cryopumps are maintained at 50--55~K.
Refilling took one hour. 
The cold switch of the He4 stage was primarily turned on
after turning off its warm switch.
The cold switch of the He3I stage was turned on when
all three temperatures of each head fell below 2~K.
The cold switch of the He3U head was turned on after a one-minute interval.
The time required to recharge was typically 3.5 hours.

\begin{figure}
 \includegraphics[width=0.84\linewidth]{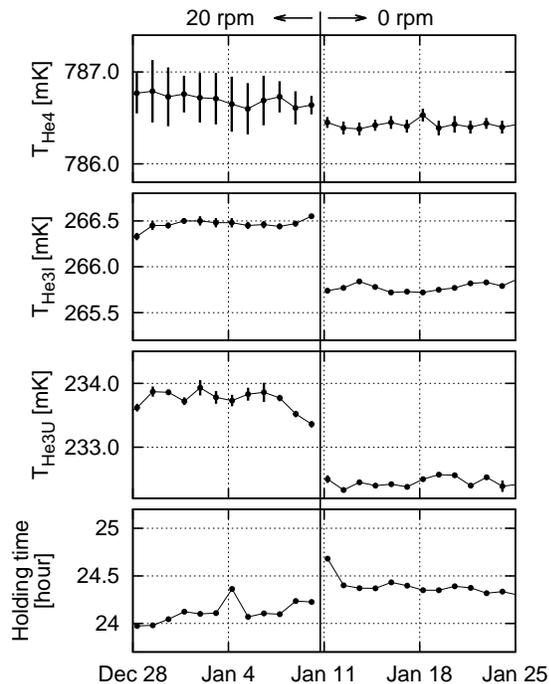}
 \caption{Four-weeks temperature trends at each head for each cycle.
 ${\rm T_{He4}}$, ${\rm T_{He3I}}$, and ${\rm T_{He3U}}$ are the minimum
 temperature at the He4, He3I, and He3U heads, respectively.
 Error bars of each data point represent fluctuations in the thermometer response.
 The bottom graph gives the trend in the holding time.
 The table was continuously rotated at 20~rpm for the first half of the period
 and stationary during the second half;
 the rotation was stopped January 10th.
 Most applications can tolerate such a few percent differences in
 performance over the two periods.}
 \label{fig:trends}
\end{figure}

Trends of the minimum temperature\footnote{
For eliminating the effects of fluctuations in thermometer response,
we took average data for a duration of 600 seconds.}
at each cold head are shown in Fig.~\ref{fig:trends}.
For the first two weeks, the table rotation was maintained at 20~rpm,
and was stopped for the second two weeks.
Error bars of each data point represent fluctuations in the thermometer response.
After the rotation was stopped, changes in the minimal temperatures
as well as the size of fluctuations indicate tiny vibrations arising
from table rotation ($< 4 \times 10^{-3}~g$ with an accelerometer)
that might cause extra thermal loading.
Heat capacity measurements of the cooler with a different setup
indicate a loading of $\lesssim$~0.1~$\mu$W for He3U. 
The extra loads also produce changes in holding times
during the interval from the time the He3U head reached a temperature
below 250~mK to the time the liquid helium in the He4 tank is exhausted.
The size of the effect is confirmed in Fig.~\ref{fig:trends}:
$\sim$~15~minutes difference over 24~hours of holding time.
The differences in temperatures and holding time were only
$\lesssim$~1\% and 2\%, respectively.
Both comply with the specifications of the cooler.
Nevertheless, most applications can tolerate such slight differences.


In summary, we developed a cryogenic system for a rotation table
that attains sub-Kelvin conditions.
The system consists of a helium sorption cooler and a pulse tube cooler.
A series of two rotary connectors for the electricity and
the helium gas circulation enables the coolers to be operated.
We performed cool-down tests on the rotating table
at the rotation speed of 20~rpm.
The catalog specifications of the coolers were confirmed;
the coldest stage achieved 0.23~K with a holding time of more than 24~hours.
The test lasted four weeks under the control of a auto-recharge scheme
(when exhausted, liquid helium was automatically refilled).
The table was continuously rotated for the first half of the period and
was stopped for the latter half.
We observed a slight difference of a few percent in performance
over the two periods.
We have established an important technology for use in various science
experiments which require sub-Kelvin temperature conditions on a
rotation platform, as for example, in the precise measurements of the
cosmic microwave background radiation using superconducting devices.


This work is supported by Grants-in-Aid for Scientific Research from The
Ministry of Education, Culture, Sports, Science, and Technology, Japan
(KAKENHI 23684017, 21111003, and 26247045).
It is also supported by the Center for the Promotion of Integrated
Sciences (CPIS) of SOKENDAI, and the Basic Science Research Program
through the National Research Foundation of Korea (NRF) funded by the
Ministry of Education, Science and Technology (2013R1A1A2004972).
We are grateful for the cooperation of Takeda Engineering Co. Ltd., and
G-tech Co. Ltd.
We also acknowledge Masaya Hasegawa, Masashi Hazumi, and Takayuki Tomaru
for their support.

%

\end{document}